\documentclass[a4paper]{article}
\usepackage{listings}
\usepackage{xcolor}
\usepackage{multirow}
\usepackage{jheppub}  
\usepackage{dcolumn}
\usepackage[english]{babel}
\usepackage[utf8x]{inputenc}
\usepackage[T1]{fontenc}
\usepackage{palatino}
\usepackage{amsmath}
\usepackage{afterpage}
\usepackage{graphicx, subcaption}
\usepackage{makeidx}
\newcommand{\be}{\begin{equation}}  
\newcommand{\ee}{\end{equation}}  
\newcommand{\bea}{\begin{eqnarray}}  
\newcommand{\eea}{\end{eqnarray}}  
\makeindex
\begin{document}

\vspace*{1.2cm}

\thispagestyle{empty}
\begin{center}
{\LARGE \bf{The FACET Project: Foward Aperture CMS ExTension\\
 to search for new Long-Lived Particles}
}
\par\vspace*{7mm}\par

{

\bigskip

\large \bf Michael G. Albrow}

\bigskip

{\large \bf  E-Mail: albrow@fnal.gov}

\bigskip

{Scientist Emeritus, Fermilab, USA}

\bigskip

{\it Presented at the Low-$x$ Workshop, Elba Island, Italy, September 27--October 1 2021}

\vspace*{15mm}

\end{center}
\vspace*{1mm}

\begin{abstract}
 FFACET is a proposed new subsysem for CMS to search for portals such as dark photons, dark higgs, heavy
neutral leptons and axion-like particles in the very forward direction at the High Luminosity LHC. 
Such particles can penetrate up to 50 m of iron and then decay inside a 14 m$^3$ vacuum pipe made by
enlarging an 18 m long section of the LHC pipe to a radius of 50 cm.
\end{abstract}

\section{Introduction}
FACET, short for \textbf{F}orward \textbf{A}perture \textbf{C}MS \textbf{E}x\textbf{T}ension,  
is a project under development to add a subsystem 
to CMS to search for beyond the standard model (BSM) long-lived particles (LLPs)
 in the high luminosity era
of the LHC, in Run 4 (2028) and beyond. 
The project was initiated with a two-day meeting in April 2020 \cite{april2020,albrowguan}, with one day discussing
a forward hadron spectrometer for strong interaction physics, and one day on searching for long-lived particles.
A description and more details of the physics potential are given in Ref. \cite{facetpaper}.

We can compare FACET to the pioneering FASER experiment \cite{faser} which is approved to search for LLPs in the very forward direction in Run 3,
and an upgrade FASER-2 \cite{faser2} which is being developed for Run 4. Major differences with FACET are (a) FASER-2 is
480 m from IR5 (with ATLAS) while FACET is 100 - 127 m from IR1 (with CMS). (b) FACET has 4$\times$ the solid
angle: 54.5 $\mu$sr cf. 13.6 $\mu$sr. (c) FASER-2 has a 5 m-long decay volume; FACET has 18 m which is evacuated to
eliminate background from particle interactions. (d) FASER-2 is centered at polar angle $\theta$ = 0$^\circ$ while
FACET covers 1 mrad $< \theta <$ 4 mrad. (e) FASER-2 is behind $\sim$ 100 m of rock absorber while FACET has $\sim$ 50 m of
iron. However FACET is located inside the main LHC tunnel where
radiation levels are much higher while FASER is located in a side tunnel. 

An important difference is that FASER is an independent experiment while FACET is not; it is proposed to be a new subsystem of CMS,
fully integrated and using the same advanced technology for its detectors. This has the added benefit of allowing
the study of correlations with the central event, and enables a standard model physics program especially in low pileup
$pp, pA$ and $AA$ collisions.

\begin{figure}[t]
  \vspace{-0.5 in}
 \begin{center}
\makebox[\textwidth][c]{\includegraphics[angle=270,origin=c,width=150mm]{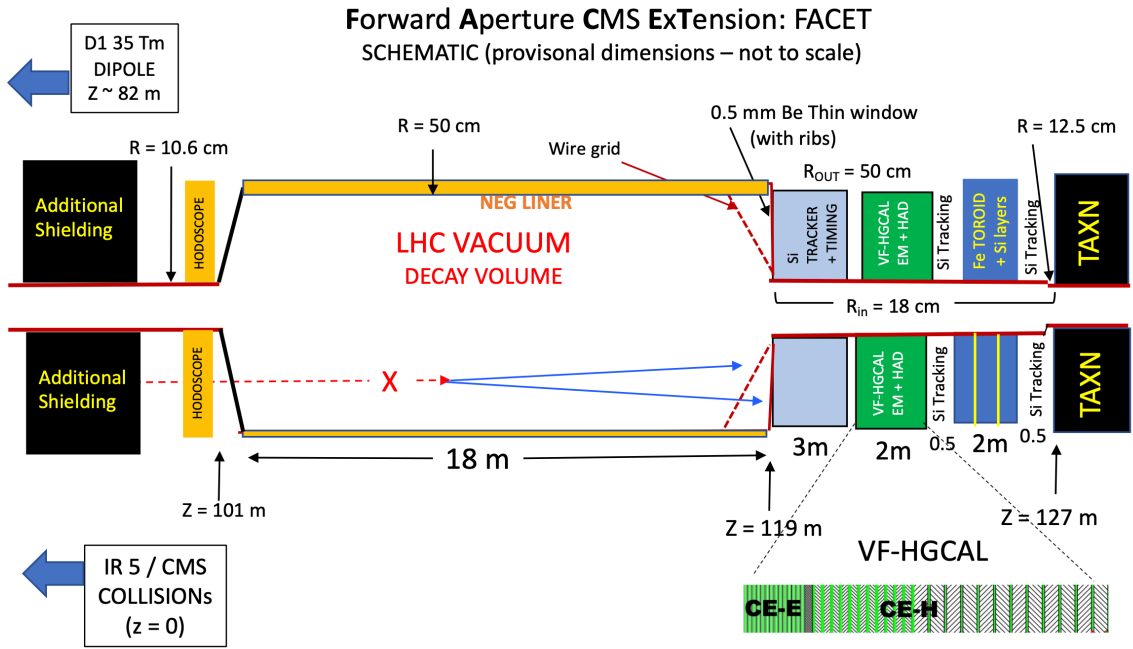}}
\end{center}
 \vspace{-1.5in}

\caption{Schematic layout of the proposed FACET spectrometer. The side view and top view are
the same since it is azimuthally symmetric. The IR5 collision region and the central CMS detectors
are 100 m to the left. An example of an LLP $X$ decaying inside the pipe is superimposed.}
\label{sketch}
 \end{figure}

 FACET will be located downstream of IR5 (at $z$ = 0) in an LHC straight section between the new (for Run 4) 
superconducting beam separation dipole D1 at $z$ = 80 m and the TAXN absorber at $z$ = 128 m. 
A schematic layout of the spectrometer is shown in Fig.~\ref{sketch}.
The beam pipe
between $z$= 101 m and 119 m will be enlarged to a radius of 50 cm. In front of the entrance window
will be a radiation-hard ``tagging'' hodoscope, with 2 - 3 planes of $\sim 1$ cm$^2$ quartz or radiation-hard scintillator
blocks. 
This must have very high efficiency with a precision time measurement for charged
particles entering the pipe. These are all background particles to be ignored in the subsequent analysis.
Excellent time resolution, $\sim$ 30 ps, together
with fast timing on the tracks from another plane between the tracker and the calorimeter 
will not only help the rejection of incoming background tracks but allows a study of their  momenta and composition.

Neutral LLPs produced with polar angle 
1 $< \theta <$ 4 mrad penetrate the iron of the LHC elements (quadrupoles Q1 - Q3  and dipole D1) and enter the big vacuum tank
where decays to SM particles can occur. The LHC-quality vacuum completely eliminates any background from interacting
particles inside a fiducial region starting behind the front window. The back window of the big pipe, where it transitions
from $R$ = 50 cm to $R$ = 18 cm, will be thin, e.g. 0.5 mm of Be with strengthening ribs,
 to minimise multiple scattering of the decay tracks\footnote{The front window may also need to be thin to minimize interactions
behind the tagging hodoscope; this is under study.} 
Behind that window, in air, the detector elements will be
3 m of silicon tracking (resolution $\sigma_x = \sigma_y \sim 30 \; \mu$m per plane) followed by a layer of
fast timing ($\sigma_t \sim$ 30 ps)\footnote{Since this Low-x Workshop 
we note that measuring the time-of-flight of these background tracks over the 22 m between the two 
hodoscopes, with a resolution $\delta \beta \lesssim 5 \times 10^{-4}$, together with the energy measured in the calorimeter,
will be very interesting. For example, consider particles with a delay relative to $\beta$ = 1 of 1 ns $\pm$ 50 ps  with a shower
of energy $E_{cal}$. These can be 0.63 GeV/c $\mu^\pm$, 0.83 GeV/c $\pi^\pm$, 3 GeV/c $K^\pm$, or 5.6 GeV/c $p$ or $\bar{p}$, easily
distinguished thereby measuring the identify and spectrum of these background tracks. That would be useful for testing and tuning
\textsc{fluka}, the LHC standard for machine protection etc. It also uniquely enables calibration of the HGCAL with hadrons of known
momenta up to tens of GeV with high  statistics even in short runs. The charge $Q$ is known from the Cherenkov light 
amplitude and track $dE/dx$
enabling measurements of light isotopes with lifetimes $\gtrsim 10^{-9}$ s in the showers, and to search for
objects such as strangelets (nuclei with extra strange quarks and therefore anomalous low charge:mass ratio).}.

The tracking and timing will be followed by a high granularity
electromagnetic and hadronic calorimeter, the HGCAL design. Muons that penetrate the HGCAL are detected in more
silicon tracking through an iron toroid. 

FACET is complementary to all other searches with unique 
access to regions of mass and coupling (or lifetime) for many \emph{portals}, hypothetical particles
that couple very weakly to both standard model particles (directly or through mixing) and to dark matter 
particles. Unlike most searches in the central detectors FACET  is sensitive to a wide variety of possible LLPs.
It has the potential to discover dark photons ($A'$), dark higgs ($h$ or $\phi$), heavy neutral 
leptons ($N_i$) and axion-like particles ($ALP$s or $a$) if they have large enough production cross section in the
very forward direction, small enough coupling to penetrate 300 $\lambda_{int}$ of iron, and lifetime
in the range $c\tau$ = 10 cm - 100 m before decaying to standard model charged particles and/or photons. 
A key feature is the high (LHC quality) vacuum tank for decays,
1 m diameter and 18 m long (14 m$^3$), made by enlarging a section of the LHC beam pipe. This allows some channels,
e.g. $X^0 \rightarrow$ multihadrons, $ \tau^+ \tau^-, c + \bar{c}$ and $b + \bar{b}$ to have zero background even in
3 ab$^{-1}$, while $e^+e^-$ and $\mu^+\mu^-$ decays may have very low backgrounds especially for
masses $\gtrsim$ 0.8 GeV. In 3 ab$^{-1}$ we expect to observe several thousand $K^0_L \rightarrow \mu^+ \mu^-$ and also
$K^0$ decays to 4 charged tracks, compromising the region around $M(X^0) =$ 0.5 GeV. 

Dark photons $A'$ are hypothetical neutral gauge bosons that do not have direct couplings with SM particles,
but they can interact indirectly by mixing with SM photons. If $M(A') < 1$ GeV their main production mechanism
is via the decays $\pi^0, \eta^0, \eta' \rightarrow \gamma \gamma$, the fluxes being highest at small polar angle $\theta$.
Fixed target experiments such as NA62 have higher luminosity, and the higher $\sqrt{s}$ of the LHC is not advantageous for
dark photons from these sources.
For  $M(A') > 1$ GeV the higher $\sqrt{s}$ of the LHC
is important, as additional sources such as Drell-Yan and quark- and proton-bremsstrahlung 
dominate. The LHC is essential if the source is a massive state such as a $Z'$ in the model of Ref.\cite{duaprime},
which would give FACET sensitivity up to $\sim$ 20 GeV.
The decay modes are the same as the final states in $e^+e^- \rightarrow \gamma^*$, with $\tau^+ \tau^-$,
$c + \bar{c}$ and multihadron decays being background-free above their thresholds. 
Measuring the relative rates of different channels could establish the identity of candidates as dark photons. 

Heavy neutral leptons $N_i$ (where $i$ represents flavor, perhaps with three different mass states to discover)
are present in many BSM theories; they may explain the light neutrino masses through the seesaw mechanism. Possible decay
modes are $N_{\mu} \rightarrow \mu^\pm W^{*\mp}$ with the virtual $W^*$ decaying to kinematically allowed leptonic
or hadronic channels, and the same modes but with $\mu^\pm$ replaced by $e^\pm, \tau^\pm$. If $N_i$ have masses in the few-GeV region
even a few good candidate events would be a discovery that would open a very rich new field of neutrino physics.
In the model of Ref.\cite{suchita} FACET has unique discovery reach up to $\sim$25 GeV.

\begin{figure}[t]
\vspace{-1.0in}
 \begin{center}
\makebox[\textwidth][c]{\includegraphics[angle=0,origin=c,width=120mm]{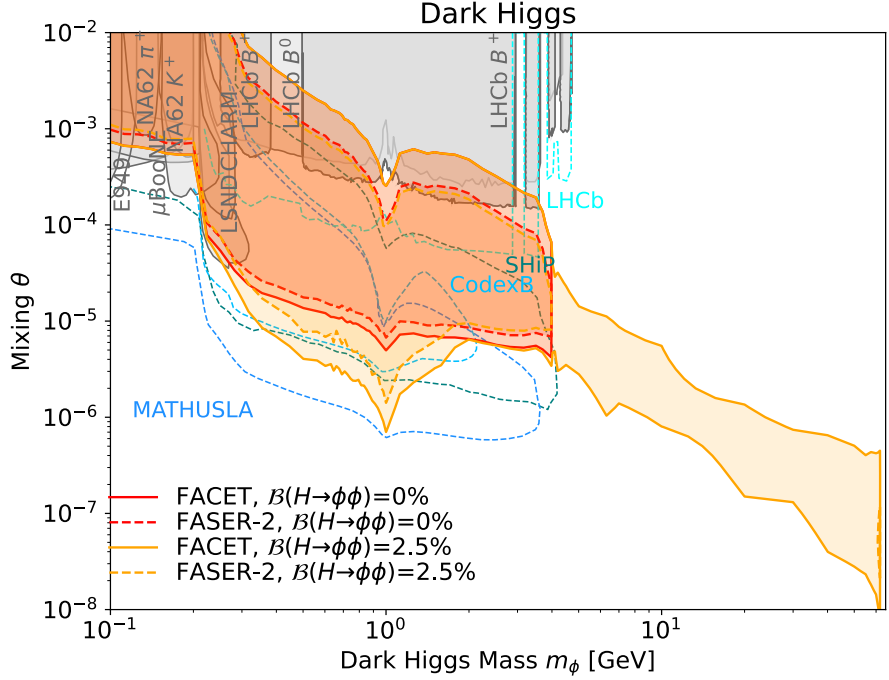}}
\end{center}
\vspace{-1.2in}
\caption{Reach of FACET and other existing and proposed experiments for a dark Higgs boson $\phi$
with the assumption of either 0\% (red lines) or 2.5\% (yellow lines) branching 
fraction for the  $H(125) \to \phi\phi$ decays. FACET offers a unique coverage all the way to half $M_H$ 
for a range of mixing angles. FACET and FASER-2 contours are calculated with \textsc{Foresee}~\protect\cite{foresee}.
Figure from Ref. \cite{facetpaper} which gives citations.}
\label{darkhiggs}
 \end{figure}

Also very exciting would be the discovery of another Higgs boson, a dark higgs, $h$ or in general a scalar $\phi$, 
having the same vacuum quantum numbers
as the H(125) but with mass possibly in the several GeV region. Present measurements of H(125) decays allow
an invisible decay fraction up to 5\%, which could be explained by an $h$ through mixing $H(125) \leftrightarrow h$ or 
decay $H(125) \rightarrow h + h$\footnote{If one $h$ is detected in FACET the other should be more central and give rise to
missing transverse energy $E_T$. Whether it is possible to detect this in a high pileup bunch crossing remains to be seen.}.
If $M(h) \lesssim$ 4.5 GeV rare $b$-decays are a potential source, with competition especially from B-factories and LHC-b. 
For 4.5 GeV $< M(h) <$ 60 GeV and a range of mixing angles FACET has unique coverage, as shown in Fig. \ref{darkhiggs}.
The most spectacular
decays are $h \rightarrow \tau^+ \tau^-, c \:+ \: \bar{c}$ and $b \: + \: \bar{b}$ if kinematically allowed, and with the heaviest
states favored; the scalar nature can be demonstrated by the relative decay fractions as well as the isotopic decay.
FACET has more sensitivity than FASER-2 due to its larger solid angle and longer decay volume, e.g.  
if there is no background and if 10 candidates were to be detected in FACET, FASER-2 would expect $<$ 1
\footnote{While the coverage of FACET is limited by LHC restrictions
in the horizontal direction, the solid angle could be increased nearly a factor $\times$2 in the vertical direction 
with a non-circular beam pipe.}.

Another possible portal is a heavy ALP, but the main decay mode to $\gamma + \gamma$ will have a high background
from random pairs of photons from $\pi^0$ and $\eta$ decay, etc. Even though the electromagnetic section of the HGCAL
measures the shower direction the vertex resolution is much worse than for charged tracks.

FACET will be live for every bunch crossing, with an expected pileup of $\sim$ 140 inelastic collisions, giving a total
integrated luminosity of $\sim$ 3 ab$^{-1}$. The \textsc{fluka} code, which is the LHC standard,  predicts about 25 charged particle tracks
with 18 cm $< R <$ 50 cm in each bunch crossing. Their origin (apart from any BSM signal!) is (a) from interactions
of beam halo and secondary particles with the beam pipe, collimators, magnets, etc. (b) from decays of neutral
hadrons, mainly $K^0_S, K^0_L$, and $\Lambda^0$. (c) Very small angle ($\theta \lesssim$ 1 mrad) charged particles that
pass through the D1 aperture, which deflects them to the left and right sides. The acceptance for the latter is 
limited to $\sim$ 2 TeV,
but they allow some standard model physics (e.g. measuring $\mu^+\mu^-$ pairs at Feynman-$x_F \sim$ 0.5).
 
In a fast Level-1 trigger the tracks will be projected upstream to the
2D  hodoscope in front of the front window. The main purpose of the hodoscope is to tag all entering charged
particles with very high efficiency (inefficiency $\lesssim 10^{-5}$) and ignore them; they are all background. 
Because of the high resolution of the tracker and because there is 
no significant magnetic field the uncertainty on the projected entrance point is $<$ 1 mm. 
Since in 3 ab$^{-1}$ there will be about $2 \times 10^{15}$ bunch crossings,
we still expect $\sim 10^5 - 10^6$ bunch crossings with two untagged tracks from different collisions entering the
decay volume, depending on the tagger inefficiency. However, the probability that these two background tracks intersect in space, i.e. have a distance of closest approach
$\lesssim 100 \; \mu$m inside the fiducial decay volume and matching in time effectively kills this pileup background.

Decays of $K^0_S, K^0_L$, and $\Lambda^0$ inside the pipe are a serious background for any LLPs with mass $M(X^0) \lesssim$
0.8 GeV decaying to hadrons. Their mass and momentum are reconstructed from the tracks and calorimeter energies (or muon momenta in the toroid),
and one can require pointing back to the IR, good timing and a flat distribution of decay distance (as it would be
for an LLP). However the background to a search for 2-body hadronic decays of an LLP is expected to be overwhelming
except for $M(X^0) \gtrsim$ 0.8 GeV. For higher masses 4-body decays become more probable, and a well-defined
vertex with $\geq$ 4 charged tracks should have zero background. The probability of two unrelated $K^0$ decays occurring
within the resolution in $x,y,z,t$ is very small but is being evaluated, as are all expected possible backgrounds. 

A Letter of Intent to CMS is being prepared to officially propose FACET as a new subsystem and initiate a technical design study.
The most critical item is the enlarged beam pipe, since that cannnot be installed in short technical 
stops, and the next planned long shutdown LS4 is in 2031. New sources of funding will be sought. 
The detectors required represent $\lesssim$ 5\% of the CMS forward upgrades, and could 
be installed (and upgraded if needed) in technical stops.

\section*{Acknowledgements}

The author thanks all the members of the FACET developing team, named as co-authors of Ref. \cite{facetpaper}. We acknowledge
valuable input from V. Baglin and P. Fessia (CERN) on the LHC pipe and infrastructure and V. Kashikhin (Fermilab) on the preliminary
toroid design. 

\pagebreak

\end{document}